%% file: paper.tex
\documentstyle[epsf]{mn}
\def\spose#1{\hbox to 0pt{#1\hss}}
\def\simgt{\mathrel{\spose{\lower 3pt\hbox{$\mathchar"218$}}
     \raise 2.0pt\hbox{$\mathchar"13E$}}}
\def\simlt{\mathrel{\spose{\lower 3pt\hbox{$\mathchar"218$}}
     \raise 2.0pt\hbox{$\mathchar"13C$}}}

\input{psfig.sty}
\title{The APM Survey for Cool Carbon Stars in the Galactic Halo - II 
The Search for Dwarf Carbon Stars. }
\author[E. J.Totten, M. J. Irwin, and P.A. Whitelock]
  {E. J. Totten,$^1$, M. J.~Irwin$^2$ and P.A. Whitelock$^{3}$\\
  $^1$Department of Physics,
	Keele University,
	 Keele, Staffordshire, ST5 5BG, UK\\
  $^2$Institute of Astronomy, Madingley Road,
	Cambridge CB3 1HA, UK\\
  $^3$South African Astronomical Observatory, P.O. Box 9, 7935 Observatory, S. Africa }

\date{Draft version}
\pubyear{1997}

\begin{document}

\maketitle

\begin{abstract}

We present proper motion measurements for carbon stars found during
the  APM Survey for Cool  Carbon Stars in  the  Galactic Halo ( Totten
$\&$ Irwin, 1998).  Measurements are  obtained using a combination  of
POSSI, POSSII and  UKST survey plates supplemented where necessary by CCD
frames taken at the Isaac Newton Telescope.  We find no significant proper 
motion for any of the new APM colour-selected carbon stars and so conclude 
that there are no dwarf carbon stars present within this sample.
We also present  proper motion measurements for three previously
known  dwarf carbon  stars  and demonstrate that these measurements  
agree favourably with those previously quoted in the literature, verifying
our method of determining proper motions.  Results from a complimentary  
program of JHK photometry obtained at the South African Astronomical 
Observatory are also presented.  Dwarf carbon stars are believed to have 
anomalous near-infrared colours, and this feature is used for further 
investigation of the nature  of the APM  carbon stars.  Our results support  
the use of  JHK photomtery as a dwarf/giant discriminator and also reinforce 
the conclusion that none of the new APM-selected carbon stars are dwarfs.  
Finally, proper motion measurements combined with extant JHK  photometry are 
presented for a sample of previously known Halo carbon stars, suggesting that 
one of these stars, CLS29, is likely to be a previously unrecognised dwarf 
carbon star.

\end{abstract}

\begin{keywords}
stars: carbon -- stars: surveys -- astrometry: stars -- infrared: stars
\end{keywords}

\section{Introduction}

Until   recently it was believed  that  all faint high-latitude carbon (FHLC)
stars were giants.  Since  main sequence stars  do not produce carbon,
it was assumed  that faint carbon stars were  distant examples  of the
classical, bright carbon giants. \\
Recently,  however a new and growing  class  of so-called dwarf carbon stars 
has been  discovered (see Green 1996 for a recent review).
Dwarf carbon stars are believed to be binary  systems, where the dwarf carbon
star has received material from a now ``invisible'' companion during the
ascent of  the companion star up the asymptotic giant branch (Dahn  
{\it et  al.} 1977).   Preliminary parallax studies indicate that dwarf
carbon stars have the luminosity of late main sequence dwarfs (Dearborn et al.
1986, Harris et al. 1998) but also mimic the overall spectral 
characteristics of carbon giants.  Dwarf carbon stars are recognisable by 
their relatively high proper motions and it has been suggested (Green {\it et al.} 1991) that they have anomalous JHK near-infrared colours. \\
Spectroscopically, the distinction between dwarf and giant carbon stars
is less obvious -- it has been suggested by Green {\it et al.}(1992) that the
spectra of dwarf carbon stars contain an  enhanced C$_{2}$ bandhead at
6191  $\AA$ which is correspondingly less pronounced in the spectra of giant 
carbon stars.\\
 
\input{pmtable}

\input{pmtable2}

The APM Survey for cool carbon stars in the Galactic Halo (Totten \& Irwin 
1998) found  48 carbon stars  with $11 \simlt R \simlt 17$; 
$B_J - R \simgt 2.4$, and at Galactic latitudes $|b| > 30^\circ$.  While there
is not enough gas  and dust present at such  high latitudes to support
recent star formation, the question of the origin of such carbon stars
is an  on-going question.  We have  suggested that these  carbon stars
provide observational evidence of recent tidal disruption of dwarf satellite
galaxies in the Galactic Halo (Totten \& Irwin 1998, Irwin \& Totten 1999).
We argue that many of the Halo carbon stars are  the tidal debris of
such merging events, and as such may  prove to be excellent probes of
the outer Halo.  However, it should also be born in mind that the current 
sample may  be contaminated by less distant, dwarf carbon  stars, which would
play no part in such a  scenario.  It is therefore essential that
every precaution is taken to  exclude dwarf  carbon stars from the putative
distant Halo carbon star sample and with that in mind we have carried out a 
series of proper motion measurements on: the APM carbon star sample; a 
selection of known dwarf carbon stars to verify the method; and a sample of 
other FHLC stars.\\

\begin{figure*}
\centerline{
\psfig{figure=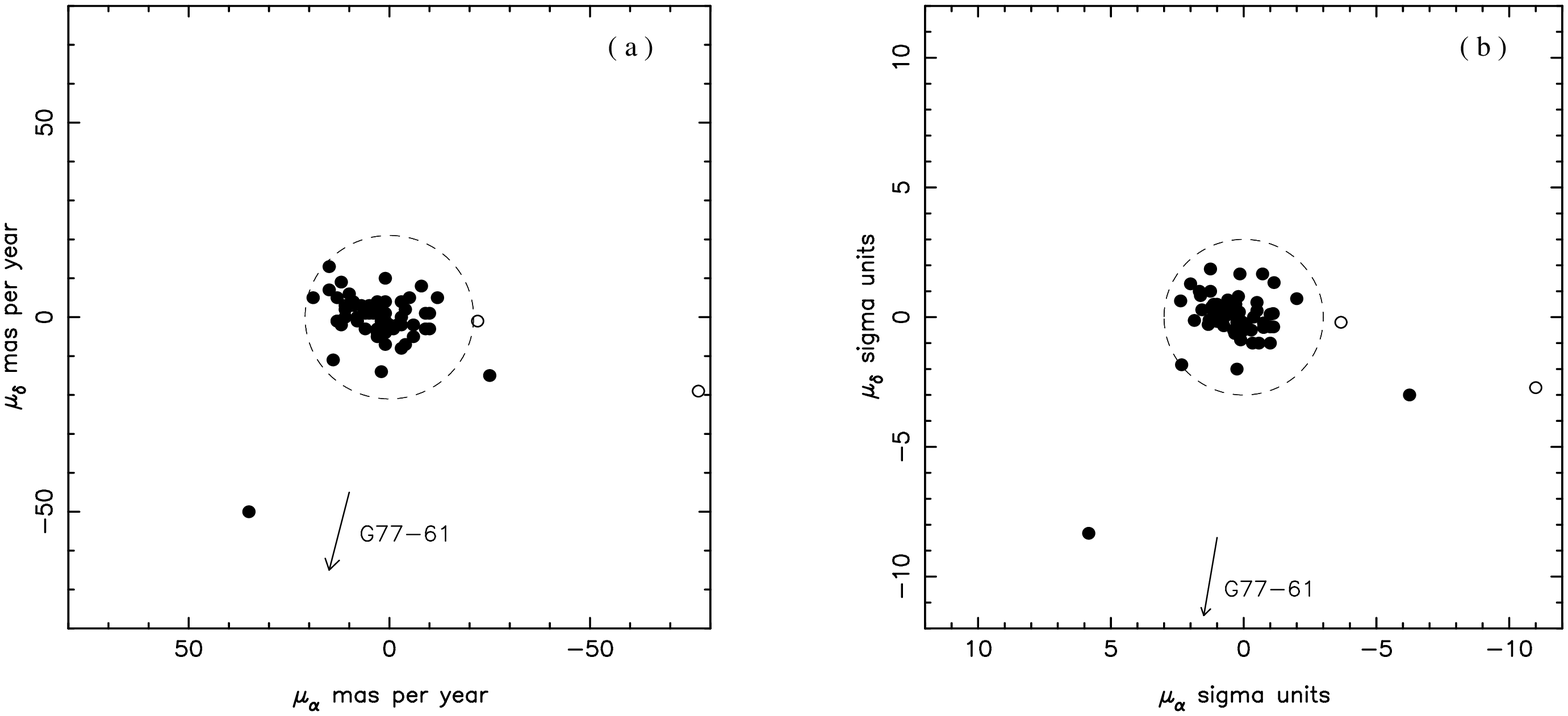,width=6.0in,height=3.0in}
}
\vspace{1.0cm}
\caption[Distribution of dwarf C stars]{(a) measured carbon star proper motions
in mas/yr.  The dashed line shows the average 3$\sigma$ limit for proper 
motion measurements (21 mas/yr).  The three dwarf carbon stars lie outside this
boundary, with G77-61 lying outside the limits of the plot. (b) the 
distribution of carbon star proper motions normalised by the error estimate for
each measurement.  The dashed line again shows the 3$\sigma$ limit for proper 
motions.  Stars outside this boundary have significant proper motions and were
subject to further investigation to determine if they are dwarf carbon stars. Dwarf carbon stars are plotted as open circles, all other carbon stars are shown as filled cirlces.}
\label{sig}
\end{figure*}

In  Section  \ref{Prop_motion}  we describe how  proper  motions were
measured for each carbon star and  show how such measurements  can readily
be used to statistically separate dwarf from giant carbon stars. Section  
~\ref{JHK} describes the  near-infrared photometry  program carried out on
the 1.9m at the South African Astronomical Observatory (SAAO) and how this  
has been applied to the classification of dwarf and giant carbon stars. Section
\ref{JHK} also includes a description  of how the JHK photometric data
can be used to estimate distances for the APM carbon stars.  \\

\section{Proper Motions}
\label{Prop_motion}

The typical tranverse velocity of Halo stars, including the effects of reflex
solar motion and the intrinsic velocity dispersion in the Halo, is 
$\approx$150 kms$^{-1}$. At a distance of 20 kpc  -- a canonical distance  
for a Halo carbon star of magnitude R $\approx$ 13.0 and with 
M$_{R} = -3.5$, this would translate to an expected proper motion of 
$\sim$ 1.5 mas$/$yr.  On the other hand, a Halo dwarf carbon star of this
apparent magnitude  and with M$_{R} \approx +9$ (Harris et al. 1998) would be 
at a distance of 
around 100 pc and would be expected to have a proper motion of $\approx$ 300 
mas$/$yr.  Even with disk kinematics the expected proper motion of a dwarf 
carbon star would be $\approx$ 100 mas$/$yr.  Clearly then, measuring proper 
motions should be a very powerful way of distinguishing between dwarf and 
giant carbon stars. \\
Suitable first epoch plate material is provided by the red glass copies (Ennnn
in Tables 1,2) of the first Palomar Sky Survey (POSSI) undertaken during the 
1950s.  For the second epoch, a mixture of the available second Palomar Sky 
Survey (POSSII) IIIaJ glass copies or IIIaF film copies (SJnnnn, SFnnnn in 
Tables 1,2), and UK Schmidt Telescope (UKST) original plates 
(Rnnnn or ORnnnn in Tables 1,2), obtained during the late 
1980s and early 1990s, provided most of the material.  In 
addition, for some of the northern hemisphere APM carbon stars we obtained 
second epoch data on the 2.5m Isaac Newton Telescope (INT) on La Palma.  The
INT data were taken in April 1997 using the first incarnation of the INT Wide 
Field Camera, with 2k $\times$ 2k thinned Loral CCDs.  The target fields were
centred on the best of the two working devices giving a field coverage
of 12.5 $\times$ 12.5 arcmin, at a spatial sampling of 0.37 arcsec/pixel.
An R-band filter was used to provide a match to the first epoch red plates
and exposure times between 50s and 100s were used to avoid saturating the
target image.\\
Much of the required plate material had already been measured on the APM 
facility (Kibblewhite {\it et al.} 1984) as part of the APM sky survey
catalogues (Irwin \& McMahon 1992, Irwin {\it et al.} 1994).  Any 
remaining second epoch POSSII plates were also measured and processed in the
same way.  The baseline difference between the two epochs for the plate
material is generally between 30--40 years, rising to well over 40 years
for the CCD data (see  Tables \ref{pm1}, \ref{pm2}). The CCD data was 
preprocessed (bias-corrected, trimmed and flat-fielded) in the standard 
manner and then an object detection and parameterisation algorithm, similar
to that run on the APM, was used to generate an object list.\\ 
Second epoch plates were also scanned around any  previously published FLHC
stars, not in the APM  survey (eg. Totten $\&$ Irwin 1998 Table 1) and of
several known dwarf carbon stars, wherever suitable plate material was readily
available.  Derived proper motions for this extra subset were then used  
to check for any previously overlooked dwarf carbon stars in the non-APM FHLC 
sample, and the published proper motions of the known dwarf carbon stars 
(eg. Deutsch 1994) were used to provide a benchmark to assess the 
external accuracy of our derived proper motions. \\

\subsection{Procedure}

The output from the measuring process, including the CCD data, is a list of 
detected objects together with a parameterisation including: x-y coordinates,
flux, and morphological information.  All plate/CCD matching was performed
in the natural plate-based coordinate system of the data.  Derived proper 
motion estimates were then transformed to a celestial system using the
usual APM plate-based astrometric calibration.   All images on both reference 
(generally second epoch) and comparison material were iteratively matched 
using a general linear 6 plate constant coordinate transform over a region of 
20 $\times$ 20 arcmin, centred on the target carbon star, for the photographic
data and 12.5 $\times$ 12.5 arcmin for the CCD data.  The linear transform
allows for shifts, rotation and scaling differences between the two epochs.
Any slight bias in the internally derived proper motion zero-point from
using all images on the plates is negligible $< 1$mas/yr compared to the
anticipated proper motion of dwarf carbon stars and also much smaller than
the random individual measuring error for each carbon star. \\
The linear transformation between 1st and 2nd epoch was derived using 
a least-squares fit  between  matched  co-ordinates of the  two  datasets.
3-$\sigma$ clipping (where  $\sigma$ = 1.48 x median absolute difference
of the coordinates) was used to reject outliers.   The least-square fits
were then repeated iteratively using 3-$\sigma$ clipping until all matched
transformed objects lie within 3-$\sigma$ of their associated positions on the 
reference system. \\
The proper motion of the target carbon  star ($\mu_{\alpha}$ and
$\mu_{\delta}$ in arcsec/yr) is then simply  the difference between the carbon
star positions on the $1^{st}$  and $2^{nd}$ epoch plates (or CCD frames), 
normalised by the epoch difference.  The typical accuracy of the positional 
difference estimate between two plates is  $\sim$ 0.27 arcsec for stellar
images and is dominated by the random errors of the individual star 
measurements.  The accuracy of the internally-derived zero-point, which
was typically based on $\approx$100 objects, is better than 1 mas/yr, again
much smaller than any expected dwarf carbon star proper motion.
For a baseline of $\sim 30-40$ years the corresponding error on the
derived carbon star proper motion is approximately 5-10 mas/yr on each
coordinate.

\subsection{ Proper Motion Results}

Tables \ref{pm1}, \ref{pm2} list the derived proper motions for the
sample studied here.  Proper motion estimates have already been
determined for the three dwarf carbon stars included in Table \ref{pm1},
by Dahn  {\it et al.} (1977), Green {\it et al.} (1991) and Heber  
{\it et al.} (1993).  Included in Table~\ref{pm1} are our estimates for 
several previously published ``normal'' FHLC stars (eg. Sanduleak \& Pesch 
1988, Bothun {\it   et al.} 1991 and  Green {\it et  al.} 1992).  Table~
\ref{pm2} lists estimates for the majority of the APM Survey carbon stars 
listed in Totten \& Irwin (1998).   The two bright objects, R$<$11, without 
a proper motion estimate in Table~\ref{pm2} are blended with neighbouring 
images on the 1st epoch POSSI plate, making it impossible to accurately 
measure the position of the carbon star.  The other two bright images are
included in the table for completeness. 

The distribution of proper motions is plotted in Figure 1. Panel 1a
shows the distribution of the  proper motions of the stars as measured,
where the  dashed line  corresponds to  the average  3-$\sigma$ error 
for  the proper motion measurements  ($\sim$ 21 mas/yr).  Any star with a 
significant proper  motion  will lie  outside  this  line.   Panel 1b  shows  
the distribution of carbon stars proper motions normalised by the individual
$\sigma$ for each measurement.   The dashed line again represents  
the notional 3-$\sigma$ boundary between a null detection of proper motion 
and stars having a significant proper motion.  The three dwarf carbon stars 
are denoted by open circles or arrows and are outside the 3-$\sigma$ boundary
on both panels.  All other carbon stars are shown as filled 
circles.  Most lie well within the null proper motion boundary and only
two fall outside of it, highlighting the simplicity of this approach for
statistically identifying dwarf proper motion candidates and also for
ruling out the dwarf hypothesis for the majority of the sample.

\subsection{The Dwarf Carbon Stars}

PG0824$+$2853 lies just outside the 3-$\sigma$ boundary in Figure 1., but
even so would have been a good candidate for a dwarf carbon star based on this
position.  Previous determinations of the proper motion reported by 
Heber et al. (1993) and by Deutsch (1994) of --0.028,0.002 and --0.036,0.000 
respectively, agree within the errors with our value of --0.022,--0.001 
arcsec/yr.  As suggested by previous authors, this relatively low proper 
motion is more consistent with a disk population dwarf carbon star, rather than
a Halo star. \\
CLS50 has a significant proper motion which places it near the edge of the
plot in  Figure 1a,b.  Green et al. (1992) first reported this object as
a dwarf carbon star and estimated its proper motion to be --0.068,--0.013
arcsec/yr.  Deutsch (1994) re-estimated the proper motion and found a result
in excellent agreement with the above of --0.069,--0.012 and in good agreement
with our derived value of --0.077,--0.019 arcsec/yr. \\
G77-61 is off the scale on both panels of the figure.
However, our value for the proper motion of 0.195,--0.745 agrees well with 
previous determinations by Dearborn et al. (1986) and by Deutsch (1994) of
0.189,--0.749 and 0.184,--0.745 respectively.  Both G77-61 and CLS50 have
the high proper motions expected from Halo population objects\\
We conclude from this brief comparison that our error estimates of between
5--10 mas/yr for the derived proper motions, based on internal statistical
considerations, provide realistic absolute proper motion errors.
\\

\subsection{Other Outliers -- A New Dwarf Carbon Star}
\label{pot-dwarf}

With the exception of two stars, all of the FHLC stars studied lie well
within  the 3-$\sigma$ boundary  and have relatively insignificant proper
motion,  leading us to conclude that none of them are dwarf carbon stars. \\

\begin{figure*}
\centerline{
\psfig{figure=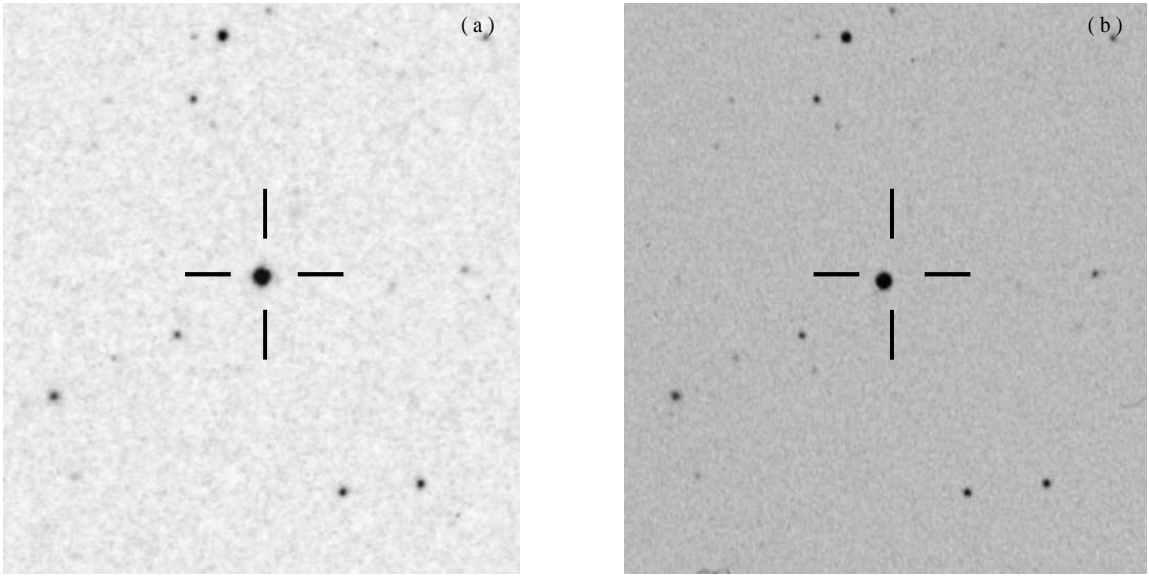,width=6.0in,height=3.0in}
}
\caption[]{5 x 5 arcmin images of the dwarf carbon star CLS29.  (a) is taken from a scan of the POSSI (first epoch) plate while (b) is taken from a scan of the POSSII (second epoch) film copy.  The images have the conventional orientation with North to the top and East to the left of the images.  The first epoch position is marked by a cross in each of the images and it is clear that the position of CLS29, relative to faint background objects, has changed  over the 35.8 year baseline between these two observations.}
\label{scans}
\end{figure*}

Of the two stars lying outside the 3-$\sigma$ boundary:
0748$+$5404 is a bright CH-type carbon  star previously reported in
Stephenson's catalogue   of  cool  carbon stars (Stephenson 1989)   and
re-observed during the APM Survey.  The brightness of this object means
that the diffraction spikes are clearly  visible on both the POSSI and POSSII
plates and that it is heavily saturated on both plates.  This leads to larger 
than average centroiding errors and  hence the
apparently  significant   proper  motion is misleading.
On further  visual examination of the  plate material by
digitally blinking pixel maps of the pair of images, we cannot detect any
obvious motion at the 1 arcsec level and therefore conclude that this star is 
probably not a dwarf carbon star.\\
CLS29$=$1037$+$3603 on the other hand is a much fainter carbon star, 
previously discovered by Sanduleak \& Pesch (1988) and re-observed during the 
APM survey.  We have detected a significant proper motion for this star of  
$\mu_\alpha = +0.035  \pm  0.006$ and  $\mu_\delta = -0.050 \pm 0.006$.  Over
a 35.8 yr baseline this corresponds to a total motion of 2.2 arcsec and should
be readily visible to the eye.  To check this an 8 $\times$ 8 arcmin region 
centred on the carbon star was digitised at 1/2 arcsec sampling for the red 
POSSI glass plate copy and the red POSSII film copy using the APM facility.
The resulting pixel maps were coaligned in a similar manner to that described
previously, using the POSSI image as reference, and using a bilinear 
interpolation to transform the POSSII image onto the reference system.
Figure~\ref{scans} shows a 5 $\times$ 5 arcmin sub-region centred on the
carbon star for each plate.  CLS29 has clearly moved relative to the majority 
of faint images, which are expected to have negligible proper motion.  Other
relatively bright images also show proper motion between the frames but this 
is to be expected given the baseline between the plates.\\
Green et al. (1992) previously investigated a sample of the then known FHLC 
stars and found CLS29 to not have a significant proper motion,  However, their
3-$\sigma$ proper motion limit of $\approx$60--70 mas/yr is a factor of
three greater than our limit of $\approx$20 mas/yr.  With a total proper
motion of 61 mas/yr (at PA 145$^\circ$), CLS29, was marginally below their
detection limit.  Such a low value for the proper motion is characteristic
of disk population dwarf carbon stars and we suspect that CLS29 may be an 
additional member of that class.

We can place additional constraints on the nature of CLS29 by combining 
the measured value for the proper motion with our published values for the
radial velocity and $B_J,R$ magnitudes (Totten \& Irwin 1998).  At Galactic 
coordinates of $l = 187.4$, $b = 60.8$, CLS29 lies within 30 degrees of the 
North Galactic pole and hence the low heliocentric radial velocity of 
$V_h = -3 \pm 6$km/s is indicative of a low value for the Z component of the 
Galactocentric velocity, consistent with disk membership.  \\
The colour transformations described in Irwin {\it et al} (1990)
and the measured photographic magnitudes for CLS29 of  $R = 14.3$ and 
$B_J = 16.4$, imply $V = 15.2$.  With the caveat of 
low number statistics, the dwarf carbon stars with measured parallaxes have a 
luminosity distribution in the range 9.6 $< M_V <$ 10.0 (Harris et al. 1998),
suggesting a distance for CLS29 of approximately 120 pc.  Combining the
proper motion, radial velocity and estimated distance yields the following
Galactocentric components of velocity, $(U,V,W) \approx (34,210,16)$, 
indicative of disk membership. \\
The spectrum of CLS29, Figure 5. of Totten \& Irwin (1998), is also
unusual.  Detailed examination of the spectrum  reveals
no obvious counterpart to the enhanced C$_2$ bandhead at 6191$\AA$, that
Green et al. (1992) have suggested is indicative of dwarf carbon stars.
Conventionally, CLS29 is a CH-type carbon star.  However, it also 
has an unambiguous H$\alpha$ emission line superimposed on the usual carbon
star spectral features.  Although not uncommon in N-type carbon stars from
our sample, CLS29 is the only CH-type carbon star we observed that has Balmer 
lines in emission.  For N-type carbon stars, Balmer emission has been 
attributed to strong chromospheric activity or Mira-like shock waves , in 
CH-type stars however, it is more likely to indicate mass 
transfer or a similar interaction between the dwarf carbon star and its  now 
invisible binary companion (eg. a faint white dwarf).  Future UV observations 
might confirm the presence of the white dwarf and hence the binary nature of 
CLS29.\\
Finally, as we show in the next section, CLS29 has JHK colours that lie close 
to the zone containing the known dwarf carbon stars with extant JHK photometry.
Taking all the evidence together strongly suggests that CLS29 is a dwarf
carbon star and that it is highly likely to be a member of the disk population
of dwarf carbon stars.

\section{ Near-Infrared JHK Photometry}
\label{JHK}

\begin{figure*}
\centerline{
\psfig{figure=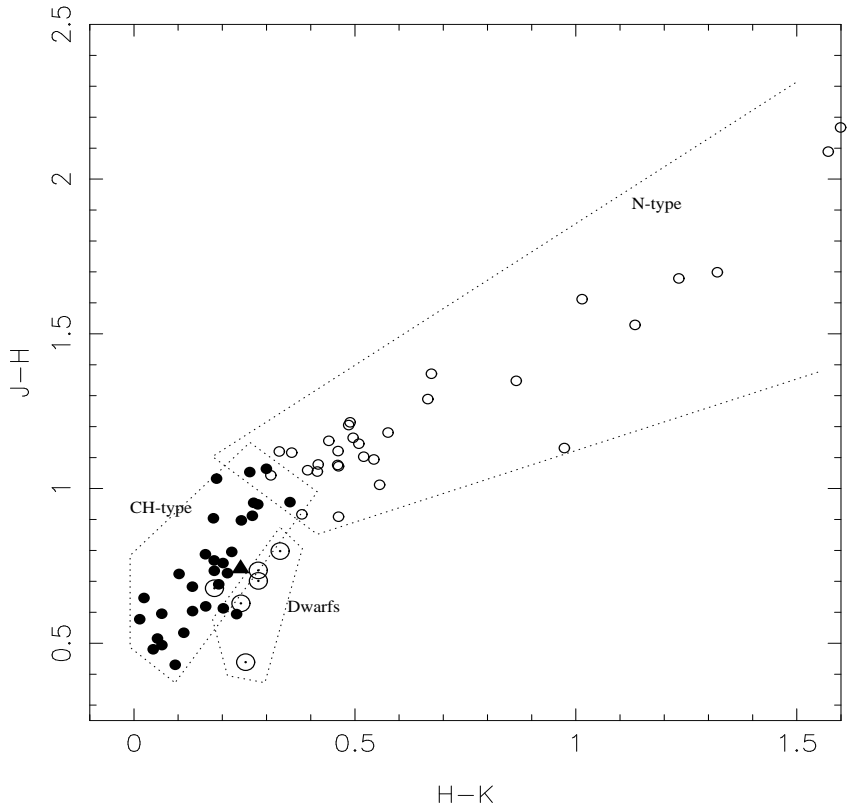,angle=0.0,width=5.0in,height=4.5in}
}
\caption[J--H vs H--K]{A two colour JHK diagram for all the FHLC stars with
extant near-infrared photometry.  Optically classified N-type carbon stars are
plotted as open circles, CH-type carbon stars are plotted as filled circles and
dwarf carbon stars are plotted as dots inside open circles.  The superimposed
dotted boundaries illustrate the locii of the separate carbon star types.
The newly discovered dwarf carbon star, CLS29$=$1037$+$3603, is plotted as a 
filled triangle and lies close to the region occupied by the known dwarf
carbon stars.}
\label{j-hh-k}
\end{figure*}

\input{apm_jhk.tex}

Green {\it et al.}(1992) and Westerlund {\it et al.}(1995) have suggested 
that dwarf carbon stars have anomalous infrared colours.  Higher gravities 
and lower metallicities encourage the association of hydrogen, thereby 
reducing the opacity in the H-band and driving the locus of dwarf carbon
star colours away from the normal carbon star locus in conventional 
two-colour JHK diagrams. 

With this in mind a program to obtain near-infrared JHK (1.25$\mu$, 1.65$\mu$,
2.2$\mu$) photometry for all the previously published carbon stars, visible 
from SAAO, was started during 1996, using the Mk III infrared photometer on 
the 1.9m telescope at SAAO.  Observations were transformed to the 
SAAO photometric system using standards taken from Carter (1990).  The mean 
JHK magnitudes for the carbon stars observed at SAAO are listed in Table 
\ref{apm-jhks}.  Associated errors for each magnitude measurement are in the 
range $\pm$ 0.01--0.02 mag for the majority of the observations, rising to
0.05--0.1 mag for those denoted by ``:''.  Several of the stars were 
observed on more than one occasion and were found to be variables.  In these
cases the average magnitude is quoted in Table~\ref{apm-jhks}.
The repeat observations, variability and periods  of such stars will be the 
subject of a forth-coming paper.  The pertinent issue here is the
luminosity class, which can be considered using the average magnitudes.

Combining the new observations with previously published JHK photometry
transformed to the SAAO system (see Carter 1990), for both the dwarf carbon 
stars and FHLC stars (see Table 1. of Totten \& Irwin 1998) yields the 
two-colour diagram in J--H -v- H--K shown in Figure \ref{j-hh-k}, where we have neglected the effects of extinction. 
All of the optically classified 
N-type stars in these tables are plotted as open circles, all CH-type stars  
as filled circles and all previously known dwarf carbon stars are  plotted 
as dots inside open circles.  As expected there is an excellent discrimination 
in colour between the locii of CH-type and N-type carbon stars, confirming
the original optical classification.  Although the dwarf carbon star population
generally lies below the locus of normal carbon stars, there is some overlap
with the distribution of normal carbon giants.  From the diagram it is 
apparent that: 
\begin{itemize}

\item  the  bluer and more common CH-type stars, mainly from the earlier FHLC 
sample, almost exclusively populate the bottom left hand corner of the  
figure.

\item  N-type carbon stars, which are also much redder optically (eg. Totten
\& Irwin 1998), are generally well separated from the CH-type stars and occupy
the upper right hand section of the diagram. 
Mira variables will fall along a general  loci similar to that of the
N-type carbon stars,  however most Miras  are extremely red and so we
would expect them to lie in the extreme top-right section of the plot,
with  some  falling outside the   boundary of the  diagram.  Both from
repeat variability measurements and from  their position in the carbon
star  locus, several  of  the N-type carbon    stars are likely to  be
carbon-rich Miras.

\item dwarf carbon stars, for which significant proper motions have been
measured, are seen to occupy a smaller section of the diagram,
along the red edge of the CH-type population.  They are unambiguously 
separated from the N-type population.

\item CLS29, the newly identified dwarf carbon star, plotted as a filled 
triangle in the figure, lies on the boundary between the CH-type and dwarf
domains.  PG0824$+$2853, which has a visible DA white dwarf companion 
possibly affecting the JHK photometry, lies inside the CH-type zone. 

\end{itemize}

So although the JHK colour alone is insufficient evidence to guarantee 
membership of the dwarf carbon star class, it does act as a useful secondary 
indicator when combined with the proper motion measurements.  JHK colour,
however, does give a good indication of generic carbon star type with a 
well-defined boundary between CH and N-type stars.  

Several of the stars in the list presented in Table 3. of Totten \& Irwin
(1998) had uncertain classification based solely on optical data. Of these:
0351$+$1127 with J--H,H--K colours of 1.12,0.33 lies just inside the N-type zone; 0936--1008 at 1.21,0.49 is well inside the N-type zone; and
1339--0700 at 0.96,0.35 lies on the boundary between CH and N-type;

Westerlund {\it et   al.}(1995) defined dwarf carbon stars as having 
J--H $<$ 0.75, H--K $>$ 0.25 mag, using the transformations of Bessell $\&$ Brett 1988, these become 0.76 and 0.26 respectively in the SAAO system, similar to that seen in Figure \ref{j-hh-k}.  This gives a clear separation between N-type and dwarf carbon  star populations, and shows independently of the proper motion measurements, that none of the N-type carbon stars found during the APM Survey are likely to be dwarf carbon stars.  CLS29$=$1037$+$3603, which has colours on the SAAO system of J--H $=$ 0.74,  H--K $=$ 0.27 lies close to the putative dwarf carbon star zone defined by
Westerlund {\it et al.}.\\

\subsection{Calibrating distances}
\label{jhkdists}

\input{dist_mod.tex}
\input{jhk_dists.tex}
\begin{figure*}
\centerline{
\psfig{figure=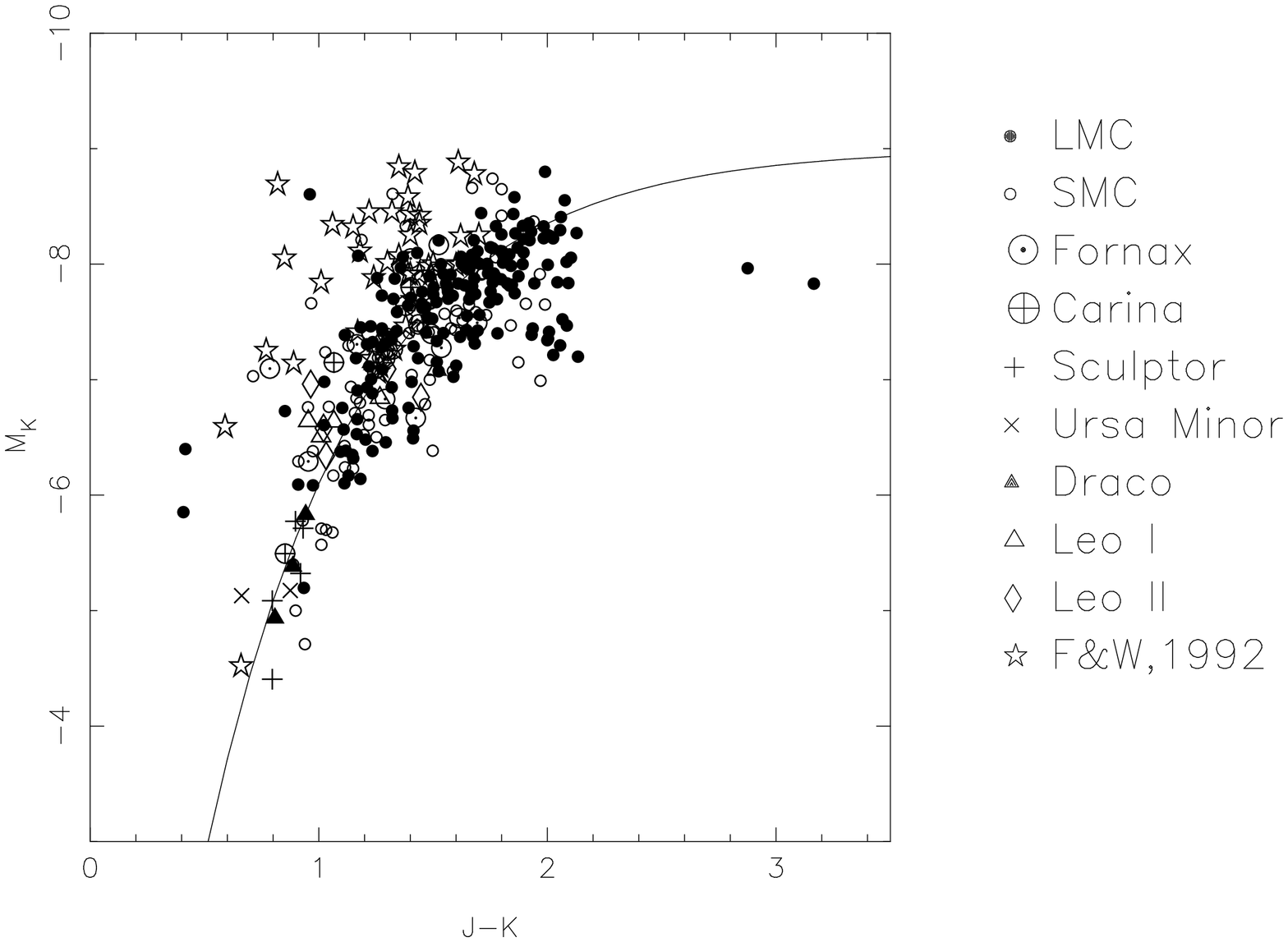,angle=-90.0,width=6.0in}
}
\caption[$M_K$ -v- J--K]{Colour Magnitude diagram showing the relationship 
between J--K  and $M_K$ for the compilation of carbon stars referred to in Table 4, including local group galaxies for which J,H,K photometry was found in the literature. The overplotted curve is an empirical fit to the data used to JHK distances to the FHLC star sample. The F$\&$W 1992 values refer to JHK measurements of the Hartwick and Cowley (1988) sample of bright LMC CH-type carbon stars. }
\label{distances}
\end{figure*}

\begin{figure*}
\centerline{
\psfig{figure=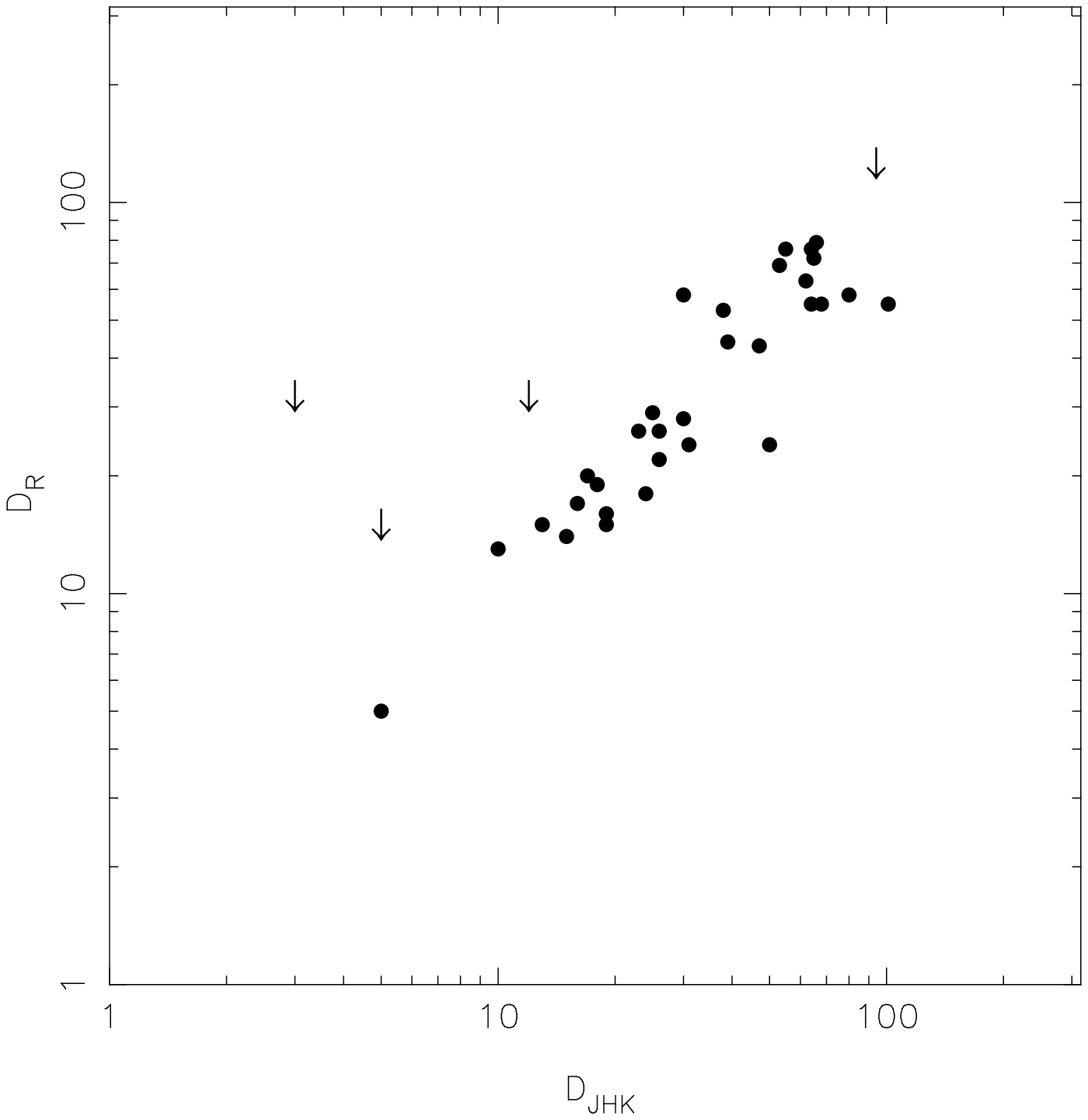,angle=-90.0,width=4.0in,height=4.0in}
}
\caption[Distance Comparison Plot.]{A comparison of distances
estimated from R magnitudes against distances estimated from JHK photometry. 
Arrows denote those stars believed to have dusty enveloping shells, for 
which the distance estimated from the R magnitude is a gross over-estimation.
This extinction will also affect the derived K-band distances but to a much
lesser extent. }
\label{jhk_dist}
\end{figure*}

Over the same period as the photometric observations of the FHLC stars
were taken, a similar set of  photometry was recorded (one dataset by MJI, 
the other by PAW $\&$ MWF) for a series of cool N-type carbon stars in the
outer parts of the SMC and LMC.  This combined dataset proved extremely 
valuable in attempting to calibrate carbon star distances derived from JHK 
photometry.

The process of calibrating stellar distances from  near-infrared photometry  
of carbon stars relies heavily on the fact that the distances to several 
nearby Galactic satellite systems are well defined and also that these dwarf 
satellite galaxies contain populations of carbon stars of similar nature to 
those found in the Halo surveys.
A compilation of carbon stars with published JHK photometry was made for
the Galactic dwarf spheroidal satellites: Fornax, Carina, Sculptor, 
Ursa Minor, Draco, Leo I and Leo II, and was combined with the aforementioned
LMC and SMC carbon star sample, supplemented by some additional LMC and SMC carbon star studies with JHK photometry.  The photometric data for the dwarf spheroidals
was taken  from  the literature.  Relevant information and references are 
given in Table  \ref{dist_moduli}.   All measurements were again transformed
to the SAAO photometric system as necessary (Glass 1985) using the equations
of Carter (1990) and Bessell and Brett (1988).  The distance moduli used were taken from a  review by Van den Bergh (1989) and are also given in Table~\ref{dist_moduli}. Extinction corrections in the K-band are less than 0.02 mag for all the carbon stars considered and we have therefore ignored the effects of extinction throughout.
Figure~\ref{distances} shows the derived absolute K-band magnitude for 
each carbon star as a function of its J--K colour.  Distance modulii for all 
stars were assumed to be the same as for the parent galaxy -- the values 
quoted in Table 4 are used in the interest of using a consistent distance 
scale, since in this context small systematic recalibration of the absolute 
distance is not of crucial importance.  In all of the following analysis we 
have also neglected the effects of intervening extinction, for both the 
calibrating carbon stars and the Halo sample, since for both samples the 
estimated JHK Galactic extinction is negligible compared to other sources of 
uncertainty.

The Hartwick and Cowley (1988) sample of bright CH-type carbon stars have
been included in Figure 4, using the JHK photometry from Feast \& Whitelock 
(1992).  (Much of the Suntzeff et al. JHK photometry for the Hartwick \& Cowley
sample is extrapolated from measurements in R and I, therefore we have used 
only the Feast \& Whitelock measurements in the figure.)  As already noted by several authors, these carbon stars appear to be significantly brighter and/or bluer than the majority of LMC carbon stars in NIR colour-magnitude diagrams.  Suntzeff {\it et al.} 1993 argued that the extant spectroscopic, photometric and kinematic data, provided a compelling case for these objects being associated with a much younger (10$^8$ years) population of AGB stars compared to the "normal" LMC carbon population (although see Costa \& Frogel 1996 for an alternative point-of-view).  
For the purposes of determining the distances listed in Table \ref{jhk_dists} we assume that the halo giants discussed here are similar
to those found in the dwarf spheroidals and to the ``normal'' carbon stars in the LMC and SMC, and hence are unlike the bright CH-type carbon stars of Hartwick $\&$ Cowley.  The possibility still remains however
that some or all of the halo CH stars might be similar to these brighter LMC carbon stars, in which case the distances listed for them in Table \ref{jhk_dists} should be regarded as lower limits.

We analysed the scatter in the distribution of carbon star JHK photometry
using Principal Components Analysis and found that the $M_K$ -v- J--K plane
produced the least scatter of any two-dimensional projection of the data
(though the average of $M_J, M_H, M_K$ was equally good).  An empirical
fit of $M_K$ with respect to J--K was made and is shown plotted as the 
solid curve in Figure~\ref{distances}.  The vertical scatter about the fitted
curve covers a range of $\approx \pm$0.5$^{m}$, with occasional more extreme 
outliers, which in the main are probably caused by variable stars.  The 
empirical fit used has the form

\begin{equation}
\log_{10}(M_K + 9.0) = 1.14 - 0.65(J-K)
\label{dist_fit}
\end{equation}

There are several caveats in using this to estimate the absolute K-band 
magnitudes, and hence distance of the Halo carbon stars.  For example: even
in the near infrared, dusty shells surrounding the central star will affect
both the J--K colour estimate and also contribute to extinction in the K-band;
we have no prior way of estimating the age of the Halo carbon star population
nor (easily) the intrinsic metallicity of the individual carbon stars, both
age and metallicity difference can cause changes in the carbon star 
luminosities; many carbon stars are known to be variables and therefore without
long term monitoring it is difficult to estimate the underlying ``true''
magnitude of this subset.  However, since the compilation of carbon stars used
to define the empirical relationship is likely to contain a similar spread of 
both age, metallicity and variable stars as the Halo population, the observed 
range in Figure~\ref{distances} of $\approx \pm$0.5$^m$provides a realistic 
error range for the distance estimate of $\approx \pm$25\%.

Equation~\ref{dist_fit} was used to estimate distances for all the FHLC star
sample with extant JHK photometry and the derived absolute K-band magnitudes, 
$M_K$  and heliocentric distances for these stars are listed in 
Table~\ref{jhk_dists}. 

A visual comparison between optical R-band distance estimates and JHK
distance estimates is shown in Figure~\ref{jhk_dist}.  
Optically-derived distances were determined for the N-type carbon stars
(Figure~\ref{jhk_dist}), by assuming a luminosity of  M$_{R}$ = --3.5
(Totten \& Irwin 1998). For the CH-type carbon stars however, the situation is
more complicated because although the optical luminosity of cool carbon stars
with optical colours of $B_J - R > 2.4$ is well approximated by a fixed 
luminosity in M$_R$, blueward of this the M$_R$ luminosity
fades rapidly with optical colour and a reliable calibration 
for the variation is not available.  Most of the wildly
discrepant points are readily explained by severe optical extinction due
to surrounding dusty shells (denoted by arrows) or by variability.  For the 
rest there is a good correlation between the optical
and JHK distance estimators for the APM sample of cool carbon stars.
One carbon star, 1225--0011, which we initially suspected from the optical
spectrum to have a dusty envelope, probably does not, given the 
agreement between the optical and near infrared distance estimates.  
The range of Halo carbon distances derived in Table ~\ref{jhk_dists}  
extends from a small number of relatively nearby ($<$10 kpc) objects, 
typically dust enshrouded, to what appear to be normal carbon star 
giants $\approx$100kpc out in the Halo of the Galaxy.

Some of the reddest carbon stars, the suspected Miras with large
amplitude magnitude variations, will have uncertain derived distances.
In particular, if the average magnitude is weighted by a severe fading
episode which can happen for some Miras, the distances will be overestimated.
More accurate distance
estimates for these objects require fuller time series analysis of the
light curves and will be presented elsewhere.

\section{Summary and Conclusions}

We have estimated proper motions for a sample of 48 carbon stars taken from
Totten \& Irwin (1998).  Because of their intrinsically faint luminosity
(Harris {\it et al.} 1998), dwarf carbon stars should have a relatively high and
measureable proper motion compared to more distant carbon giants found in the
Halo.  With a 3-$\sigma$ detection threshold of 21 mas/yr, our proper motion 
estimates show that none of the APM colour-selected carbon stars have proper 
motions indicative of dwarf carbon stars.   A complimentary program of near 
infrared JHK photometry was used as further verification of carbon star type. 
It was found that none of the APM carbon stars lie in the region of a J--H -v-
H--K plot expected to be populated by dwarf carbon stars (Westerlund 
{\it et al.} 1995 ) and this is taken as further proof that the sample of APM 
carbon stars discussed here and in Totten \& Irwin (1998) are all distant 
carbon giants.\\

As a further test, proper motion measurements and extant near infrared 
photometry were also analysed from a previously published sample of FHLC stars
and known dwarf carbon stars (Totten \& Irwin 1998 Table 1.)
with the purpose of verifying the existing luminosity classes of these stars
and checking our proper motion measurements.  It was found that one of 
these stars, CLS29$=$1037$+$3603 has a significant proper motion 
($\mu_{\alpha}$ = 0.035 $\pm$ 0.006 and $\mu_{\delta}$ = -0.050 $\pm$ 0.006 
mas/yr), indicating that it is a previously unrecognised dwarf carbon star.
We have shown that CLS29 lies within the region occupied by dwarf carbon stars
in a J--H, H--K two-colour diagram, confirming the implication of the
proper motion measurement.  CLS29 therefore appears to be an additional member
of the handful of currently known dwarf carbon stars. Further examination of 
the space motion of CLS29, making plausible assumptions regarding the intrinsic
luminosity, show that it has a Galactocentric velocity of (U,V,W) $\approx$ 
(34, 210, 16) suggesting it is highly likely to be a dwarf carbon star 
belonging to the Galactic disk population.\\

Finally, by compiling a list of JHK photometry for known carbon stars in
Galactic satellite dwarfs and supplementing it with SAAO JHK photometry 
for a large sample of cool carbon stars in the outer halos of the SMC and LMC,
we have produced a calibration linking M$_K$ and J--K for the FHLC sample.  
An empirical fit to this distribution was used to determine absolute K-band 
magnitudes for all of the FHLC stars with available JHK photometry.  These 
absolute magnitude estimates were then used to derive approximate distances to
the Halo sample of carbon stars accurate to a range of $\pm$25\%, equivalent
to a $\sigma_{dist} \approx$0.15\%.  A comparison of distances estimated from 
JHK observations and distances deduced from optical observations, for the 
mainly N-type cooler carbon stars, shows a good correlation for most of
the sample.  For a small subset of N-type carbon stars, the effect of 
extinction from a putative dusty enveloping shell is clearly seen in the 
differences between the JHK and the R-band derived distances, and is also 
apparent in the published optical spectra.\\

The original primary aims of the APM carbon star survey were: to find a 
well-defined sample of FHLC stars covering all of the high Galactic latitude 
sky; to prove that the majority, if not all, of these carbon stars are giants 
and hence members of the Halo; and to aquire accurate radial velocities 
together with good distance estimates for the entire sample, via JHK 
photometry.  The survey is now complete over 3/4 of the high latitude sky; 
none of the newly discovered carbon stars have been found to be dwarf carbon 
stars; and we have accurate radial velocities and JHK photometric distances, 
for the majority of the sample.  Interpreting the phase 
space structure seen in this distant Halo carbon star sample and using it
as a probe of the Galactic Halo will be the subject of future papers.

\section{Acknowledgments}

We would like to take this opportunity to thank M.W. Feast for helpful comments
and for contributing to the JHK observing progam, and the
Director of SAAO for the use of SAAO facilities.
Thanks are also due to the UKSTU for providing the plate material used
in the southern part of the survey, and to members of the APM facility, past
and present for maintaining such an excellent system.  The CCD 
observations were made on the Isaac Newton Telescope which is operated on 
the island of La Palma by the Isaac Newton Group in the Spanish Observatorio 
del Roque de los Muchachos of the Instituto de Astrofisica de Canarias.
This research has made use of the Simbad database, operated at CDS, 
Strasbourg, France.

\end{document}

%% file: pmtable.tex
\begin{table*}
\begin{minipage}{150mm}
\begin{center}
\label{pm1}
\caption{Measured proper motions for a selection of dwarf carbon stars and 
previously published faint high latitude carbon stars. }
\vspace{0.5cm}
\begin{tabular}{|l||l|llccc|c}
\hline
\multicolumn{1}{|c|}{\sf Name} &
\multicolumn{1}{|c|}{\sf Coord} &
\multicolumn{1}{ l}{\sf $1^{st}$ Epoch} & 
\multicolumn{1}{ l}{\sf $2^{nd}$ Epoch} & 
\multicolumn{1}{|c|}{\sf Baseline } &
\multicolumn{1}{ c}{\sf $\mu_{\alpha}$ } &
\multicolumn{1}{ c}{\sf $\mu_{\delta}$ } &
\multicolumn{1}{ c}{\sf Notes }  \\
\multicolumn{1}{|c|}{\sf } &
\multicolumn{1}{|c|}{\sf (B1950)} &
\multicolumn{1}{ l}{\sf (POSSI)} & 
\multicolumn{1}{ l}{\sf (POSSII/UKST)} & 
\multicolumn{1}{|c|}{\sf (yr)} &
\multicolumn{1}{ c}{\sf (arcsec/yr)} &
\multicolumn{1}{ c}{\sf (arcsec/yr)}  &
\multicolumn{1}{ c}{\sf  }  \\
\hline
{\bf Dwarf}& & & & & \\
G77$-$61 & 0330$+$0148 & E932  & R11438  & 32.9 & $+$0.199 $\pm$ 0.006 & $-$0.740 $\pm$ 0.007  & 1  \\
         &   & E932  & SJ01677 & 34.1 & $+$0.191 $\pm$ 0.006 & $-$0.750 $\pm$ 0.006  &\\
PG0824   &0824$+$2853  & E1351 & SJ03130 & 35.0 & $-$0.022 $\pm$ 0.006 & $-$0.001 $\pm$ 0.005  & 2  \\
CLS50 & 1217$+$3704 & E1599 & SJ01744 & 31.9 & $-$0.077 $\pm$ 0.007 & $-$0.019 $\pm$ 0.007  & 3  \\
\hline
{\bf FHLC}& & & & &  &\\
C*01&0002$+$0053 & E319  & R7939   & 30.9 & $-$0.010 $\pm$ 0.009 & $+$0.001 $\pm$ 0.007  & 4 \\
C*03&0108$-$0015 & E1259 & OR11269 & 31.8 & $-$0.004 $\pm$ 0.008 & $+$0.002 $\pm$ 0.008  & 4\\
&            & E1259 & SJ03587 & 35.9 & $-$0.006 $\pm$ 0.008 & $-$0.002 $\pm$ 0.009  &\\
Green et al.&0311$+$0733 & E1471 & SJ03564 & 34.8 & $+$0.000 $\pm$ 0.004 & $-$0.003 $\pm$ 0.006  & 5 \\
CLS26&1025$+$2923 & E1387 & SJ01033 & 31.8 & $+$0.009 $\pm$ 0.008 & $+$0.004 $\pm$ 0.008  & 6  \\
CLS43&1135$+$3321 & E109  & SJ01730 & 37.8 & $-$0.003 $\pm$ 0.008 & $+$0.000 $\pm$ 0.007  & 6 \\ 
CLS45&1149$+$3727 & E109  & SJ01730 & 37.8 & $+$0.002 $\pm$ 0.007 & $+$0.000 $\pm$ 0.007  & 6 \\
CLS54&1231$+$3625 & E105  & SJ01744 & 37.8 & $+$0.001 $\pm$ 0.009 & $-$0.007 $\pm$ 0.008  & 6 \\
CLS57&1239$+$3122 & E64   & SJ03131 & 39.9 & $-$0.009 $\pm$ 0.008 & $-$0.003 $\pm$ 0.008  & 6 \\
CLS80&1415$+$2936 & E70   & SJ04486 & 42.0 & $+$0.002 $\pm$ 0.008 & $-$0.014 $\pm$ 0.007  & 6 \\
CLS98&1605$+$2922 & E134  & SJ03297 & 40.0 & $-$0.008 $\pm$ 0.007 & $+$0.008 $\pm$ 0.006  & 6 \\
CLS112&1706$+$3316 & E1132 & SJ01967 & 34.0 & $+$0.014 $\pm$ 0.006 & $-$0.011 $\pm$ 0.006  & 6 \\
C*02&2354$+$0021 & E319  & R7939   & 31.0 & $+$0.001 $\pm$ 0.009 & $-$0.001 $\pm$ 0.007  & 7 \\
\hline
\end{tabular}
\end{center}
Notes : 1  Dahn {\it et al.} 1977; 2   Heber {\it et al.} 1993; 3  Green {\it et al.} 1992; 4  Bothun {\it et al.} 1991; 5 Green {\it et al.} 1994; 6  Sanduleak $\&$ Pesch 1988; 7  Bothun {\it et al.} 1991.   
\end{minipage}
\end{table*}

%% file: pmtable2.tex
\begin{table*}
\begin{minipage}{150mm}
\caption{Measured Proper Motions for APM colour-selected faint high latitude
carbon stars. }
\begin{center}
\label{pm2}
\vspace{0.5cm}
\begin{tabular} {|lllc|cc|l|}
\hline
\multicolumn{1}{|c|}{\sf Name} &
\multicolumn{1}{ l}{\sf $1^{st}$ Epoch} & 
\multicolumn{1}{ l}{\sf $2^{nd}$ Epoch} & 
\multicolumn{1}{|c|}{\sf Baseline } &
\multicolumn{1}{ c}{\sf $\mu_{\alpha}$ (arcsec/yr)} &
\multicolumn{1}{ c}{\sf $\mu_{\delta}$ (arcsec/yr)} \\
\hline
0000$+$3021 & E1257 & SJ05402 & 38.9 & $-$0.002 $\pm$ 0.007 & $-$0.002 $\pm$ 0.004  \\
0102$-$0556 & E1206 & OR13284 & 35.0 & $+$0.006 $\pm$ 0.008 & $-$0.003 $\pm$ 0.009  \\
0123$+$1233 & E635  & SJ01389 & 34.7 & $-$0.003 $\pm$ 0.006 & $+$0.004 $\pm$ 0.007  \\
0207$-$0211 & E825  & R8222   & 29.1 & $+$0.009 $\pm$ 0.009 & $+$0.003 $\pm$ 0.007  \\
0217$+$0056 & E1283 & R8284   & 29.0 & $-$0.003 $\pm$ 0.009 & $-$0.008 $\pm$ 0.008  \\
0222$-$1337 $^3$ & E886  & R7240   & 27.9 & $+$0.015 $\pm$ 0.012 & $+$0.013 $\pm$ 0.007  \\
0225$+$2634 & E858  & SJ04231 & 37.9 & $+$0.008 $\pm$ 0.008 & $-$0.001 $\pm$ 0.006  \\
0316$+$1006 & E11   & SF04359 & 42.0 & $+$0.001 $\pm$ 0.005 & $-$0.004 $\pm$ 0.006  \\
0340$+$0701 & E1499 & SF03652 & 34.9 & $-$0.006 $\pm$ 0.006 & $-$0.005 $\pm$ 0.005  \\ 
0351$+$1127 & E940  & SF02252 & 34.9 & $+$0.000 $\pm$ 0.006 & $-$0.002 $\pm$ 0.007  \\
0357$+$0908 & E1308 & SF02252 & 34.0 & $+$0.001 $\pm$ 0.004 & $-$0.001 $\pm$ 0.005  \\ 
0418$+$0122 & E1524 & OR14085 & 35.4 & $+$0.001 $\pm$ 0.005 & $+$0.004 $\pm$ 0.005  \\
0713$+$5016 & E670  & SF02976 & 36.8 & $-$0.003 $\pm$ 0.004 & $-$0.002 $\pm$ 0.005  \\
0748$+$5404 $^3$ & E985  & SF06135 & 41.0 & $-$0.025 $\pm$ 0.004 & $-$0.015 $\pm$ 0.005  \\ 
0748$+$7221 $^3$ & E680 $^{1}$  &  CCD    & 44.1 &      &      \\
0911$+$3341 & E1342 &  CCD    & 42.1 & $+$0.012 $\pm$ 0.006 & $+$0.009 $\pm$ 0.007  \\
0915$-$0327 & E430  & R8400   & 31.2 & $+$0.010 $\pm$ 0.008 & $+$0.006 $\pm$ 0.006  \\
0936$-$1008 & E1536 & OR11832 & 31.4 & $+$0.001 $\pm$ 0.006 & $+$0.001 $\pm$ 0.005  \\
0939$+$3630 & E925  &  CCD    & 43.3 & $-$0.005 $\pm$ 0.007 & $+$0.005 $\pm$ 0.003  \\
1013$+$7340 & E685  &  CCD    & 44.1 & $+$0.004 $\pm$ 0.004 & $+$0.001 $\pm$ 0.002  \\
1019$-$1136 & E1537 & OR9900  & 29.1 & $+$0.015 $\pm$ 0.009 & $+$0.007 $\pm$ 0.007  \\
1037$+$3603 & E731  & SF02321 & 35.8 & $+$0.035 $\pm$ 0.006 & $-$0.050 $\pm$ 0.006  \\
1037$+$2616 & E1380 &  CCD    & 42.0 & $+$0.003 $\pm$ 0.005 & $+$0.002 $\pm$ 0.003  \\
1056$+$4000 & E1349 & SJ03815 & 35.9 & $+$0.005 $\pm$ 0.009 & $+$0.003 $\pm$ 0.008  \\
1123$+$3723 & E695  & SJ03164 & 37.1 & $-$0.004 $\pm$ 0.007 & $-$0.007 $\pm$ 0.007  \\
1130$-$1020 & E1562 & OR12389 & 31.9 & $+$0.001 $\pm$ 0.007 & $+$0.010 $\pm$ 0.006  \\
1211$-$0844 & E1023 & OR10243 & 31.2 & $+$0.013 $\pm$ 0.008 & $+$0.005 $\pm$ 0.006  \\
1225$-$0011 & E1405 & OR12996 & 33.8 & $+$0.008 $\pm$ 0.011 & $+$0.000 $\pm$ 0.007  \\
1241$+$0237 & E104  & R5032   & 29.1 & $-$0.009 $\pm$ 0.009 & $+$0.001 $\pm$ 0.009  \\ 
            & E104  & SJ03232 & 40.0 & $-$0.010 $\pm$ 0.010 & $-$0.003 $\pm$ 0.008  \\
1249$+$0146 & E1578 & R5032   & 23.3 & $+$0.011 $\pm$ 0.009 & $+$0.003 $\pm$ 0.009  \\
            & E1578 & SJ03232 & 34.2 & $+$0.002 $\pm$ 0.007 & $+$0.000 $\pm$ 0.010  \\
1254$-$1130 & E1591 & OR13095 & 33.1 & $+$0.013 $\pm$ 0.010 & $-$0.001 $\pm$ 0.009  \\
1339$-$0700 & E500  & OR9964  & 32.8 & $+$0.019 $\pm$ 0.008 & $+$0.005 $\pm$ 0.008  \\
1350$+$0101 & E465  & R6808   & 29.1 & $+$0.007 $\pm$ 0.010 & $+$0.003 $\pm$ 0.009  \\
1406$+$0520 & E96   &  CCD    & 47.0 & $+$0.012 $\pm$ 0.009 & $-$0.002 $\pm$ 0.007  \\
1429$-$0518 & E1062 & R9158   & 29.9 & $+$0.013 $\pm$ 0.007 & $-$0.001 $\pm$ 0.008  \\
1442$-$0058 & E1613 & R5774   & 22.9 & $-$0.001 $\pm$ 0.011 & $-$0.003 $\pm$ 0.010  \\
1450$-$1300 & E1025 & OR11939 & 32.8 & $+$0.006 $\pm$ 0.007 & $+$0.001 $\pm$ 0.006  \\
1509$-$0902 & E1431 & OR14962 & 37.0 & $+$0.003 $\pm$ 0.008 & $-$0.003 $\pm$ 0.006  \\
1511$-$0342 & E1431 & R6979   & 26.0 & $+$0.003 $\pm$ 0.010 & $+$0.004 $\pm$ 0.008  \\
1519$-$0614 & E1431 & R6979   & 26.0 & $+$0.003 $\pm$ 0.010 & $-$0.004 $\pm$ 0.009  \\
1523$+$4235 & E1376 & SF05116 & 38.0 & $-$0.012 $\pm$ 0.006 & $+$0.005 $\pm$ 0.007  \\
2111$+$0010 & E575  & OR13307 & 37.1 & $+$0.006 $\pm$ 0.005 & $+$0.002 $\pm$ 0.005  \\ 
	    & E575  & SJ03390 & 37.9 & $+$0.001 $\pm$ 0.007 & $+$0.001 $\pm$ 0.008  \\
2213$-$0017 & E1146 & OR11414 & 32.2 & $+$0.005 $\pm$ 0.008 & $+$0.001 $\pm$ 0.008  \\ 
            & E1146 & SJ02069 & 34.0 & $+$0.002 $\pm$ 0.011 & $-$0.001 $\pm$ 0.006  \\
2223$+$2548 & E817  & SJ02069 & 34.1 & $+$0.003 $\pm$ 0.009 & $-$0.005 $\pm$ 0.008  \\
2225$-$1401 & E1180 & OR13265 & 35.0 & $+$0.011 $\pm$ 0.010 & $+$0.000 $\pm$ 0.008  \\
2229$+$1902 & E842  & SJ03396 & 36.8 & $+$0.011 $\pm$ 0.007 & $+$0.002 $\pm$ 0.007  \\
2255$+$0556 & E821  & SJ03416 & 36.9 & $+$0.000 $\pm$ 0.008 & $-$0.002 $\pm$ 0.009  \\ 
2259$+$1249 $^3$ & E800 $^{2}$ & SF04266 & 38.1 &      &     \\   
\hline
\end{tabular}
\end{center}
Notes : \\
1.  Image is elliptical and unresolved on plate E680.\\
2.  Image is elliptical and unresolved on plate E800.\\
3.  These four Carbon stars brighter than the APM survey limit: 0222$-$1337; 
0748$+$5404; 0748$+$7221; 2259$+$1249 have been included here for 
completeness.

\end{minipage}
\end{table*}

%% file: apm_jhk.tex
\begin{table*}
\begin{minipage}{150mm}
\caption[Infrared photometry for carbon stars.]{Infrared photometry of 
FLHC stars obtained at SAAO}
\label{apm-jhks}
\begin{center}
\begin{tabular} {lcccrl}
\hline
\multicolumn{1}{c}{\sf Name} & 
\multicolumn{1}{ c}{\sf J} & 
\multicolumn{1}{ c}{\sf H} & 
\multicolumn{1}{ c}{\sf K} &
\multicolumn{1}{ c}{\sf  } & 
\multicolumn{1}{ l}{\sf Comments}  \\ 

\hline
0002$+$0053  & 11.11  & 10.35  & 10.10  & C*01 CH-type & \\
0100$-$1619  & 13.14: & 12.59: & 12.51: & C*23 CH-type & \\
0102$-$0556  & 11.34  &  9.74  &  8.60  & C*07 $+$ APM N-type& $+$em lines, variable -- Mira ? \\
0108$-$0015  & 13.55: & 12.84: & 12.77:  & C*03 CH-type & \\
0123$+$1233  &  9.85  &  8.89  &  8.62   & Stephenson $+$ APM CH-type & \\
0207$-$0211  & 11.25  & 10.33  &  9.90   & C*30 $+$ APM N-type & variable -- Mira ? \\
0217$+$0056  & 11.34  & 10.23  &  9.82   & C*31 $+$ APM N-type & \\
0222$-$1337  &  7.45  &  6.38  &  6.09   & C*15 CH-type & IRAS FSC \\
0225$+$2634  & 10.50  &  9.28  &  8.79   & APM N-type   & \\
0228$-$0256  & 12.57  & 11.81  & 11.63   & C*08 CH-type & \\
0229$-$0316  & 11.95  & 11.02  & 10.69   & C*09 CH-type & \\
0311$+$0733  & 14.01: & 13.032 & 12.87:  & Green et al. 1994 CH-type& \\ 
0316$+$1006  & 10.51  &  9.62  &  9.37   & APM CH-type  & \\
0340$+$0701  &  9.97  &  8.81  &  8.37   & APM N-type   & \\
0351$+$1127  & 11.82  & 10.70  & 10.37   & APM N-type   & \\
0357$+$0908  & 12.79  & 11.61  & 11.26   & APM N-type   & \\
0418$+$0122  & 10.07  &  7.98  &  6.41   & APM N-type   &$+$dust -- Mira ? IRAS PSC2.0 \\
0915$-$0327  &  9.97  &  8.91  &  8.65   & APM CH-type  & \\
0936$-$1080  & 12.41  & 11.20  & 10.72   & APM N-type   & \\
1019$-$1136  & 10.09  &  9.00  &  8.45   & APM N-type   & \\
1025$+$2923  & 12.29  & 11.70  & 11.64:  & CLS26 CH-type& \\
1037$+$2616  & 11.34  & 10.28  &  9.88   & APM N-type   & \\
1130$-$1020  &  7.39  &  5.23  &  3.63   &IRAS $+$ APM N-type&$+$dust -- Mira ?\\
1211$-$0844  & 13.19  & 12.28  & 11.82   & APM N-type & $+$em lines -- Mira ? \\
1220$+$2122  & 11.59  & 10.45  & 9.94    & Moody et al. $+$APM N-type& \\
1225$-$0011  & 12.27  & 11.13  & 10.16   & APM N-type   & -- Mira ? \\
1241$+$0237  & 12.93  & 11.86  & 11.39   & UM515 $+$ APM N-type & \\
1249$+$0146  & 12.76  & 11.42  & 10.55   & APM N-type   & -- Mira ? \\
1252$+$1017  & 12.16  & 11.48  & 11.31   & CLS67 CH-type & \\
1254$-$1130  & 10.94  &  9.89  &  9.47   & APM N-type   & \\
1339$-$0700  & 10.52  &  9.56  &  9.21   & APM CH/N-type& \\
1350$+$0101  & 12.73  & 11.55  & 10.97   & APM N-type   & \\
1406$+$0520  &  8.84  &  7.74  &  7.22   & Stephenson $+$ APM N-type \\
1415$+$2936  & 10.82  & 10.33  & 10.17   & CLS80 CH-type & \\
1429$-$0518  & 13.94: & 12.26  & 11.03   & APM N-type   &$+$dust -- Mira ?  \\
1442$-$0058  & 12.22  & 11.14  & 10.68   & APM N-type   & \\
1450$-$1300  & 13.90: & 12.29  & 11.27   & APM N-type   & -- Mira ? \\
1511$-$0342  & 12.83: & 11.82  & 11.26   & APM N-type   & \\
1519$-$0614  & 13.26  & 12.09  & 11.60   & APM N-type   & \\
1525$+$2912  & 13.60: & 12.80  & 12.76:  & CLS87 CH-type & \\
1605$+$2922  & 10.78  & 10.13  & 10.01   & CLS98 CH-type & \\
2111$+$0010  & 12.60  & 11.70  & 11.52   & APM CH-type  & \\
2213$-$0017  & 10.91  &  9.83  &  9.41   & APM N-type   & \\
2213$-$1451  & 12.13  & 11.22  & 11.03   & C*26 CH-type & \\
2223$+$2548  &  7.65  &  5.95  &  4.63   & Stephenson $+$ APM N-type &$+$dust, em lines -- Mira ? IRAS PSC2.0  \\
2225$-$1401  & 12.13  & 10.76  & 10.09   & APM N-type   & variable -- Mira ?  \\
2229$+$1902  & 10.89  &  9.77  &  9.31   & APM N-type   &  \\
2255$+$0556  & 10.63  &  9.71  &  9.45   & APM CH-type  &  \\
2259$+$1249  &  7.11  &  5.83  &  5.16   & Stephenson $+$ APM N-type& $+$em lines  \\
2305$-$1356  & 14.38: & 13.75: & 13.80:  & C*13 CH-type & \\
2346$+$0248  & 13.14: & 12.40: & 12.26:  & C*10 CH-type & \\
2354$+$0021  & 13.46  & 13.09: & 12.81:   & C*02 CH-type & \\
\hline
\end{tabular}
\end{center}
Note : See Tables 1 and 3, Totten $\&$ Irwin 1998 for further details.\\  
\end{minipage}
\end{table*}

%% file: dist_mod.tex
\begin{table}
\caption{Adopted parameters for the satellite galaxies of the Milky Way used
in constructing Figure~\ref{distances}.}
\label{dist_moduli}
\begin{tabular} {lccr}
\hline
\multicolumn{1}{l}{\sf Galaxy} & 
\multicolumn{1}{ c}{\sf Distance modulus$^{*}$} &
\multicolumn{1}{ c}{\sf E(B-V) } \\ 
\hline
Carina $^{1}$     &  20.14 & 0.025$^{a}$\\  
Draco  $^{2}$     &  19.4  & 0.03 $^{a}$ \\
Fornax $^{3}$     &  20.59 & 0.03 $^{b}$ \\
Leo I  $^{2}$     &  21.8  & 0.00 $^{a}$ \\
Leo II $^{2}$     &  21.85 & 0.02 $^{b}$ \\
LMC    $^{4}$ &  18.45  & 0.074$^{b}$ &  \\
Sculptor $^{5}$   &  19.47 & 0.02 $^{a}$  \\
SMC    $^{4}$ &  18.80  & 0.03 $^{a}$    \\
Ursa Minor$^{2}$  &  19.3  & 0.02 $^{a}$  \\
\hline
\end{tabular}
\\
\\
Notes: \\
$^{*}$ \hspace{0.75cm} taken from Van den Bergh 1989.\\
$^{a}$ \hspace{0.75cm} E(B-V) taken from Aaronson \& Mould 1985.\\
$^{b}$ \hspace{0.75cm} E(B-V) taken from Feast \& Whitelock 1992.\\
$^{1}$ \hspace{0.75cm} photometry from  Mould {\it et al.} 1982.\\
$^{2}$ \hspace{0.75cm} photometry from  Aaronson $\&$ Mould 1985. \\
$^{3}$ \hspace{0.75cm} photometry from  Aaronson $\&$ Mould 1980. \\
$^{4}$ \hspace{0.75cm} unpublished photometry from  Irwin, Feast $\&$ Whitelock. \\
$^{ }$ \hspace{0.90cm} photometry from Westerlund {\it et al.} 1995.\\
$^{ }$ \hspace{0.90cm} photometry from Feast $\&$ Whitelock 1992.\\
$^{5}$ \hspace{0.75cm} photometry from  Frogel {\it et al.} 1982  . \\

\end{table}

%% file: jhk_dists.tex
\begin{table*}
\begin{center}
\vspace{0.5cm}
\caption[Photometric distances in kpc for all previously published carbon stars.]
{Photometric distances in kpc for all carbon stars with extant infrared photometry.}
\label{jhk_dists}
\vspace{0.5cm}
\begin{tabular}{lcccccccc}
\hline
\multicolumn{1}{c}{\sf Name} & 
\multicolumn{1}{ c}{\sf $M_{K}$} &
\multicolumn{1}{ c}{\sf d$_{\odot}$(JHK)} & &
\multicolumn{1}{c}{\sf Name} & 
\multicolumn{1}{ c}{\sf $M_{k}$} &
\multicolumn{1}{ c}{\sf d$_{\odot}$(JHK) } \\ 
\hline
0002$+$0053  &   --6.1 &   18   && 1231$+$3625  &   --5.9  &    9   \\ 
0100$-$1619  &   --3.9 &   20   && 1239$+$3122  &   --5.2  &    7   \\    
0102$-$0556  &   --8.8 &   30   && 1241$+$0237  &   --7.7  &   66   \\	
0108$-$0015  &   --5.0 &   35   && 1249$+$0146  &   --8.5  &   65   \\ 
0123$+$1233  &   --6.9 &   13   && 1252$+$1017  &   --5.4  &   22   \\
0207$-$0211  &   --7.3 &   27   && 1254$-$1130  &   --7.6  &   26   \\ 
0217$+$0056  &   --7.7 &   31   && 1313$+$3535  &   --5.5  &   25   \\
0222$-$1337  &   --7.3 &    5   && 1339$-$0700  &   --7.2  &   19   \\   
0225$+$2634  &   --8.0 &   23   && 1350$+$0101  &   --8.1  &   64   \\
0228$-$0256  &   --5.8 &   31   && 1406$+$0520  &   --7.9  &   10   \\    
0229$-$0316  &   --7.0 &   35   && 1415$+$2936  &   --4.1  &    7   \\
0311$+$0733  &   --6.6 &   80   && 1429$-$0518  &   --8.8  &   94   \\   
0316$+$1006  &   --6.6 &   16   && 1442$-$0058  &   --7.7  &   47   \\ 
0340$+$0701  &   --7.8 &   17   && 1450$-$1300  &   --8.7  &  101   \\
0351$+$1127  &   --7.5 &   38   && 1509$-$0902  &   --5.8  &  117   \\
0357$+$0908  &   --7.7 &   62   && 1511$-$0342  &   --7.8  &   64   \\
0418$+$0122  &   --8.9 &   12   && 1519$-$0614  &   --7.9  &   80   \\ 
0911$+$3341  &   --4.2 &    9   && 1523$+$4235  &   --8.6  &   20   \\   
0915$-$0327  &   --7.2 &   15   && 1525$+$2912  &   --5.3  &   41   \\ 
0936$-$1080  &   --8.0 &   55   && 1532$+$3242  &   --5.9  &   25   \\
0939$+$3630  &   --3.6 &   21   && 1605$+$2922  &   --4.9  &    9   \\
1015$+$3540  &   --5.2 &   26   && 1637$+$3437  &   --4.1  &   16   \\   
1019$-$1136  &   --7.9 &   19   && 1706$+$3316  &   --3.1  &   15   \\ 
1025$+$2923  &   --4.1 &   14   && 2111$+$0010  &   --6.4  &   39   \\     
1037$+$2616  &   --7.5 &   30   && 2213$-$0017  &   --7.6  &   25   \\ 
1037$+$3603  &         &        && 2223$+$2548  &   --8.9  &    5   \\     
1123$+$3723  &   --6.0 &   11   && 2225$-$1401  &   --8.4  &   50   \\ 
1130$-$1020  &   --9.0 &   3    && 2229$+$1902  &   --7.8  &   26   \\  
1135$+$3321  &   --5.0 &   32   && 2255$+$0556  &   --6.8  &   18   \\   
1149$+$3727  &   --5.2 &   13   && 2259$+$1249  &   --8.3  &    5   \\  
1211$-$0844  &   --7.3 &   68   && 2305$-$1356  &   --7.8  &   35   \\
1213$+$3721  &   --3.1 &    9   && 2346$+$0248  &   --5.5  &   36   \\     
1220$+$2122  &   --7.9 &   37   && 2354$+$0021  &   --4.1  &   24   \\ 
1225$-$0011  &   --8.4 &   53   &&              &          &        \\
\hline
\end{tabular}
\end{center}
\vskip 1.0truecm
\end{table*}